\documentclass[prd,aps,superscriptaddress,floatfix,nofootinbib,eqsecnum,twocolumn]{revtex4-2}

\pdfoutput=1

\usepackage{amsmath}
\usepackage{amsfonts}
\usepackage{amsmath}
\usepackage{amssymb}
\usepackage{bm}
\usepackage{dcolumn}
\usepackage{graphicx}
\usepackage[utf8]{inputenc}
\usepackage{latexsym}
\usepackage{rotating}
\usepackage{hyperref}
\usepackage{subfigure}
\usepackage{color}
\usepackage{subcaption} 
\usepackage{graphicx}   
\usepackage{changes}
\usepackage{verbatim}
\usepackage{comment}
\usepackage{empheq}
\usepackage{csquotes}
\usepackage{physics}
\usepackage{xcolor}
\usepackage{float}
\usepackage{soul}
\usepackage{orcidlink}
\usepackage{amsmath,latexsym}
\usepackage{booktabs}  

\begin{document}
\title{Unified Framework for Geodesic Dynamics with Conservative, Dissipative, and GUP Effects}

\author{Gaurav Bhandari\orcidlink{0009-0001-1797-2821}}\email{bhandarigaurav1408@gmail.com, gauravbhandari23@lpu.in.}\affiliation{Department of Physics, Lovely Professional University, Phagwara, Punjab, 144411, India.}

\author{S. D. Pathak\orcidlink{0000-0001-9689-7577}}\email{sdpathak@lko.amity.edu; prince.pathak19@gmail.com}\affiliation{Amity School of Applied Sciences, Amity University Uttar Pradesh, Lucknow Campus, Lucknow, 226028, India.}

\author{Harjit  Singh Ghotra \orcidlink{0000-0001-6380-7563}}\email{hs10_phy@yahoo.co.in}\affiliation{Department of Physics, Lovely Professional University, Phagwara, Punjab, 144411, India}

\author{Maxim Yu Khlopov \orcidlink{0000-0002-1653-6964}}\email{khlopov@apc.in2p3.fr}\affiliation{Virtual Institute of Astroparticle Physics, 75018 Paris, France.}
\author{Maxim A. Krasnov }\email{morrowindman1@mail.ru}\affiliation{Institute of Physics, Southern Federal University, Rostov on Don 344090, Russia}
\affiliation{National Research Nuclear University MEPHI, 115409 Moscow, Russia.}

\begin{abstract}
We derive generalized geodesic equations in curved spacetime that include conservative forces, dissipative effects, and quantum-gravity-motivated minimal-length corrections. Conservative interactions are incorporated through external vector potentials, while dissipative dynamics arise from an exponential rescaling of the particle Lagrangian. Phenomenological study of Quantum-gravity effects is introduced via Generalized Uncertainty Principle (GUP) deformed Poisson brackets in the Hamiltonian framework. We show that free-particle geodesics remain unaffected at leading order, but external potentials induce velocity-dependent corrections, implying possible violations of the equivalence principle. As an application, we analyze modified trajectories in Friedmann-Lemaitre-Robertson-Walker (FLRW) universes dominated by dust, radiation, stiff matter, and dark energy. Our results establish a unified approach to conservative, dissipative, and GUP-corrected geodesics, providing a framework to probe the interplay between external forces, spacetime curvature, and Planck-scale physics.
\end{abstract}

\maketitle



\section{introduction}
Einstein's general theory of relativity (GR) opened new and profound directions in modern physics. Since gravity dominates on cosmic scales, GR provides a solid geometric framework for building cosmological models. Over the past century, it has successfully explained many astrophysical and cosmological phenomena, such as the anomalous perihelion shift of Mercury \cite{R1}, the bending of light by massive objects \cite{R2}, the gravitational redshift of photons \cite{R3}, and the prediction and later confirmation of black holes \cite{R4}.

General Relativity (GR) is built on two key ideas. The first is the equivalence principle, which says that gravity acts the same on all objects, no matter their shape or composition. The second is Mach's principle, which states that matter shapes the geometry of spacetime, and this curvature then guides matter to move along the straightest possible paths, called geodesics and they are crucial for understanding phenomena such as cosmic horizons, dark energy, and gravitational lensing, as well as for studying observational effects like cosmic bubble collisions and theoretical frameworks such as the AdS/CFT correspondence \cite{R5,R6,R7,R8,B11,B12,B13}.

Obtaining exact geodesic solutions is of particular significance, as they serve as a rigorous tool for testing FLRW cosmological models and for systematically investigating the affect of curvature, quintessence, and local shear on observational outcomes.

However, a free particle in a given manifold follows geodesic motion determined solely by the spacetime geometry, the presence of external forces beyond gravitation alters these geodesic trajectories. Such deviations from pure geodesic trajectories occur ubiquitously in realistic astrophysical and cosmological contexts. Conservative forces, exemplified by electromagnetic interactions, or dissipative, as in cosmological settings where the expansion of the universe induces a Hubble-like friction term \cite{R8a,R8b,R9,R10,R11}. Recent work has demonstrated that thermal friction mechanisms can simultaneously address both the Hubble tension and large-scale structure anomalies \cite{R8b}, suggesting that dissipative effects may play a fundamental role in cosmic evolution. These friction terms emerge from the coupling between scalar fields and the expanding spacetime background in the quintessence inflationary scenario \cite{R11}.

The role of geodesics in region of strong curvature and high energies where the classical description of spacetime itself becomes insufficient, requiring a full theory of quantum gravity \cite{R12,R13,R14,R15,R16,R17}. The study of geodesics and geodesic distances is thus crucial, not only for understanding classical relativistic motion but also as a probe of quantum gravitational effects in regimes where GR ceases to be valid. Although none of the quantum theories of gravity is yet complete, several leading candidates, most notably string/M-theory \cite{str1,str2,str3,str4,str5}, loop quantum gravity (LQG) \cite{lqg1,lqg2,lqg3,lqg4,lqg5}, and related models \cite{dsr1,dsr2,dsr3} share the common feature of predicting a minimal measurable length \cite{bhandari, bhandari1, gup1,gup2,gup3,gup4,gup5,gup6,gup7,gup8,gup9,gup10}. This implies the existence of a fundamental cutoff in the spacetime continuum, often interpreted as a discretization of spacetime at the Planck scale. 

Debates on the fate of the equivalence principle (EP) in quantum gravity can be traced back to the 1970s, notably sparked by the Colella-Overhauser-Werner (COW) experiment \cite{COW}, which probed the quantum interference of neutrons in Earth's gravitational field. Since then, various approaches to QG incorporating minimal length have suggested possible violations or deformations of the EP \cite{er1,new,As,tack,equv2,equv3}. Understanding whether such violations are generic features of quantum gravity or artifacts of specific models remains an open question.  

In this work, we investigate the influence of both conservative and dissipative forces on geodesic motion, and extend the analysis to quantum-gravity-motivated corrections. In particular, we explore phenomenological modifications induced by the Generalized Uncertainty Principle (GUP) \cite{gup11,gup12,gup13,gup14,bh1,bh2,bh5,bh6,bh7,R18,R19}, which incorporates a minimal length scale into the dynamics. In particular, we analyze the modifications to geodesic equations for both free particles and particles subject to external potentials.

The structure of the paper is as follows. In Sec. (\ref{sec1}), we derive the geodesic equation in the presence of conservative external forces using a variational approach, and subsequently extend the analysis to include dissipative forces.We examine the modified geodesic equations in FLRW spacetime and illustrate their qualitative behavior in Sec.~(\ref{sec2}). In Sec. (\ref{sec3}), we study modifications to the geodesic equation from the Hamiltonian formulation incorporating a minimal length scale, considering both the free particle Hamiltonian and the Hamiltonian of a particle subjected to a potential. Finally, Sec. (\ref{sec4}) summarizes our results and outlines possible directions for future work.

\section{Geodesic equation}\label{sec1}
In differential geometry and general relativity, a  geodesic represents the extremal path between two spacetime events, generalizing the notion of a straight line in flat (Euclidean) space to curved manifolds. In Euclidean geometry, the geodesic corresponds to the shortest distance between two points, i.e., a straight line. However, in a curved spacetime background, geodesics are defined as the curves that extremize the spacetime interval, they represent the trajectory of a free-falling particle under the influence of gravity alone. Consequently, any test particle that is not subject to non-gravitational (external) forces will follow a geodesic determined solely by the background metric.

When external forces are present, the motion of the particle deviates from the geodesic, and the trajectory is governed by modified equations that incorporate the additional interactions. These deviations can be understood as perturbations to the geodesic equation or derived from an action principle that includes interaction terms. There exist several approaches to derive the geodesic equations, such as extremizing the spacetime interval, applying the principle of least action, or employing the Euler--Lagrange formalism. For a particle moving between two points \( P_1 \) and \( P_2 \) in a background spacetime with potential external interactions, the action is typically written as
\begin{equation}
S = \int_{P_1}^{P_2} \mathcal{L}(x^\mu, \dot{x}^\mu, \lambda) \, d\lambda,
\end{equation}
where \( \lambda \) is an affine parameter along the trajectory and \( \mathcal{L} \) encodes both the metric and interaction structure. In the absence of external forces, the Lagrangian reduces to \( \mathcal{L} = \sqrt{-g_{\mu\nu} \dot{x}^\mu \dot{x}^\nu} \), whose variation yields the standard geodesic equation.
\subsection{Geodesics in the Presence of an External Force}\label{sec1a}
To incorporate an external force into the dynamics of a point particle, we begin with the action
\begin{equation}
\mathcal{S} = \int \left( \frac{1}{2}\, g_{\mu\nu}(x)\, \dot{x}^\mu \dot{x}^\nu 
+ V_\mu(x)\, \dot{x}^\mu \right) d\lambda,
\end{equation}
which yields the Lagrangian,
\begin{equation}
\mathcal{L} = \frac{1}{2}\, g_{\mu\nu}(x)\, \dot{x}^\mu \dot{x}^\nu 
+ V_\mu(x)\, \dot{x}^\mu,
\end{equation}
where \( g_{\mu\nu}(x) \) is the spacetime metric and \( V_\mu(x) \) is a covariant vector potential associated with a conservative external force. Introducing the Euler-Lagrange equation via variational principle,
$
\frac{d}{d\lambda}\left(\frac{\partial \mathcal{L}}{\partial \dot{x}^\mu}\right) 
- \frac{\partial \mathcal{L}}{\partial x^\mu} = 0$,
we compute
\begin{align}
\frac{d}{d\lambda}\frac{\partial \mathcal{L}}{\partial \dot{x}^\mu} 
&= \partial_\sigma g_{\mu\nu}\, \dot{x}^\nu \dot{x}^\sigma 
+ g_{\mu\nu}\, \ddot{x}^\nu
+ \partial_\sigma V_\mu \, \dot{x}^\sigma, \label{e1}\\[0.5em]
\frac{\partial \mathcal{L}}{\partial x^\mu} 
&= \tfrac{1}{2}\, \partial_\mu g_{\alpha\beta}\, \dot{x}^\alpha \dot{x}^\beta
+ \partial_\mu V_\nu \, \dot{x}^\nu. \label{e2}
\end{align}
Substituting Eqs.~\eqref{e1} and \eqref{e2} into the Euler-Lagrange equation, and using the definition of the Christoffel symbols
\[
\Gamma^\rho_{\sigma\nu} 
= \tfrac{1}{2} g^{\rho\mu} 
\left( \partial_\sigma g_{\nu\mu} + \partial_\nu g_{\sigma\mu} - \partial_\mu g_{\sigma\nu} \right),
\]
we arrive at the modified geodesic equation as
\begin{equation}\label{g1}
\frac{d^2x^\rho}{d\lambda^2} 
+ \Gamma^\rho_{\sigma\nu}\, \dot{x}^\sigma \dot{x}^\nu
= - g^{\rho\mu} \bigl( \partial_\sigma V_\mu - \partial_\mu V_\sigma \bigr) \dot{x}^\sigma.
\end{equation}
the appearance of a non-zero term on the right-hand side of Eq.~\eqref{g1} reflects the presence of an external force. In the absence of such a force, the term vanishes and the particle follows a geodesic determined solely by the background geometry. The additional contribution is antisymmetric in its indices and resembles a field-strength tensor. Indeed, when \( V_\mu \) is identified with the electromagnetic four-potential, the force term reduces to the Lorentz force law in curved spacetime. More generally, Eq.~\eqref{g1} shows that a broad class of conservative interactions can be incorporated into the geodesic framework, effectively extending the notion of geodesic motion beyond purely gravitational dynamics.

\subsection{Dissipative Geodesic Equation} \label{sec1b}

To account for dissipative effects in the motion of a point particle, we introduce a modified Lagrangian of the form \cite{R20,R21,R22}  
\begin{equation}
\mathcal{L} = e^{\Gamma(g_{\alpha\beta},x^\alpha)} 
\left[ \frac{1}{2}\, g_{\mu\nu}\, \dot{x}^\mu \dot{x}^\nu \right],
\end{equation}
where $\Gamma(g_{\alpha\beta},x^\alpha)$ is a dissipation exponent, defined as a scalar function of the spacetime metric and the coordinates. This exponential factor effectively rescales the standard point-particle Lagrangian and introduces non-conservative dynamics. 

A specific realization of the dissipation exponent is given by  
\begin{equation}\label{dissiex}
\Gamma(g_{\alpha\beta},x^\alpha) = \int \nabla_\kappa u^\kappa \, Q(x^\alpha) \, d\lambda,
\end{equation}
where $u^\kappa$ denotes the four-velocity field of the background and $Q(x^\alpha)$ is a scalar function that characterizes the strength of the dissipation. Equation \eqref{dissiex} makes explicit that dissipation arises from the divergence of the velocity field, weighted by the local interaction strength $Q(x^\alpha)$.  

The terms in the Euler-Lagrange equations involves derivative of $\mathcal{L}$ with respect to $\dot{x}^\rho$, yielding  
\begin{align}\label{de1}
\frac{d}{d\lambda}\!\left(\frac{\partial \mathcal{L}}{\partial \dot x^\rho}\right)
&= e^{\Gamma}\Bigl(
   \partial_\sigma g_{\rho\nu}\,\dot x^\sigma \dot x^\nu
   + g_{\rho\nu}\,\ddot x^\nu \notag\\
&\qquad\qquad\quad
   + (\partial_\sigma \Gamma)\,\dot x^\sigma \, g_{\rho\nu}\dot x^\nu
   \Bigr),
\end{align}
and the derivative with respect to $x^\rho$ give
\begin{equation}\label{de2}
\frac{\partial \mathcal{L}}{\partial x^\rho}
= e^{\Gamma}\left(
   \tfrac{1}{2}\,\partial_\rho g_{\alpha\beta}\,\dot x^\alpha\dot x^\beta
   + \tfrac{1}{2}\,(\partial_\rho \Gamma)\, g_{\alpha\beta}\,\dot x^\alpha \dot x^\beta
   \right).
\end{equation} 
Using Eqns.(\ref{de1},\ref{de2}), identity for Christoffel symbols and the Euler-Lagrange equation, we obtain the modified geodesic equation as 
\begin{align}\label{dgeq}
\frac{d^2x^\mu}{d\lambda^2} 
+ \Gamma^\mu_{\;\sigma\nu}\frac{dx^\sigma}{d\lambda}\frac{dx^\nu}{d\lambda} 
= \frac{1}{2} g^{\mu\rho} (\partial_\rho \Gamma)\, g_{\alpha\beta} \frac{dx^\alpha}{d\lambda}\frac{dx^\beta}{d\lambda} 
\nonumber \\  
- (\partial_\sigma \Gamma) \frac{dx^\sigma}{d\lambda}\frac{dx^\mu}{d\lambda},
\end{align}
the right-hand side of Eq.(\ref{dgeq}) encodes the deviation from geodesic motion due to dissipation. The first term corresponds to a force-like contribution proportional to the gradient of the dissipation function $\Gamma$, while the second term acts as a drag term aligned with the particle's trajectory. In the limit where the exponential dissipation term $e^\Gamma =1$ for $\Gamma = 0$, both terms vanish, and the standard geodesic equation is recovered.

\section{Application}\label{sec2}
To investigate the qualitative behaviour of the modified geodesic equation, we consider the FLRW spacetime in the presence of both conservative external forces and dissipative effects, as discussed in the following subsections.
\subsection{For Friedmann-Lemaitre-Robertson-Walker spacetime}\label{sec2a}

\begin{figure*}[h]
\centering
\begin{tabular}{ccc}
\includegraphics[width=0.27\textwidth]{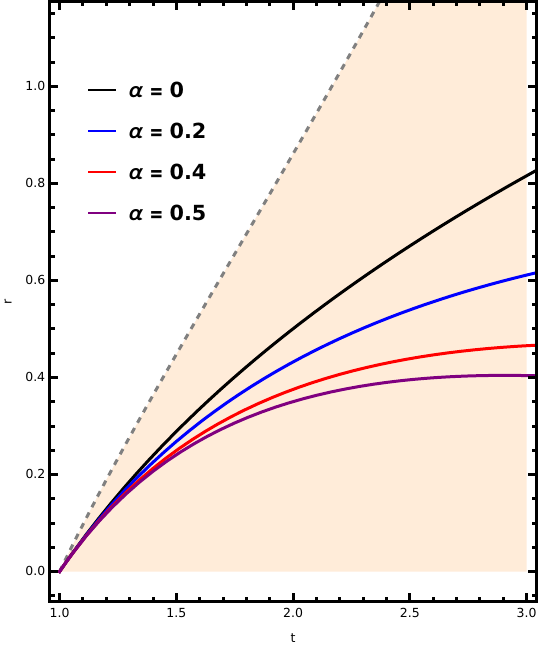} &
\includegraphics[width=0.27\textwidth]{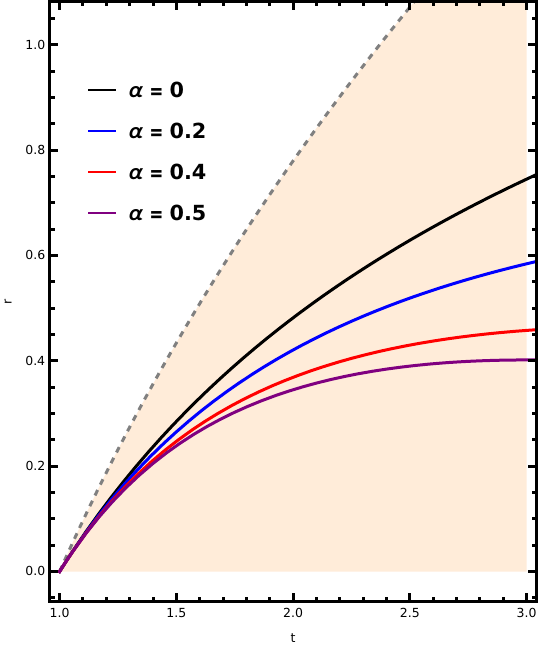} &
\includegraphics[width=0.27\textwidth]{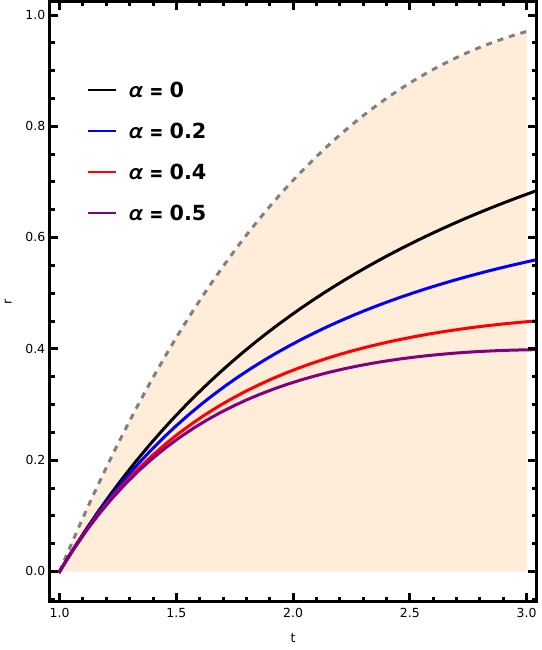} \\
\includegraphics[width=0.27\textwidth]{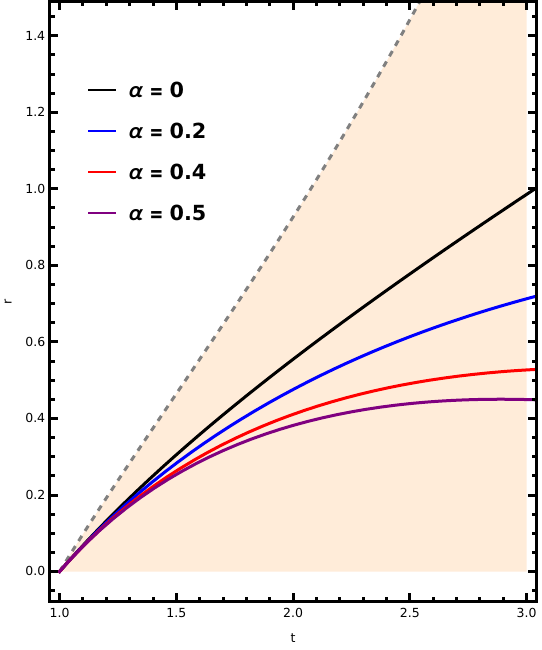} &
\includegraphics[width=0.27\textwidth]{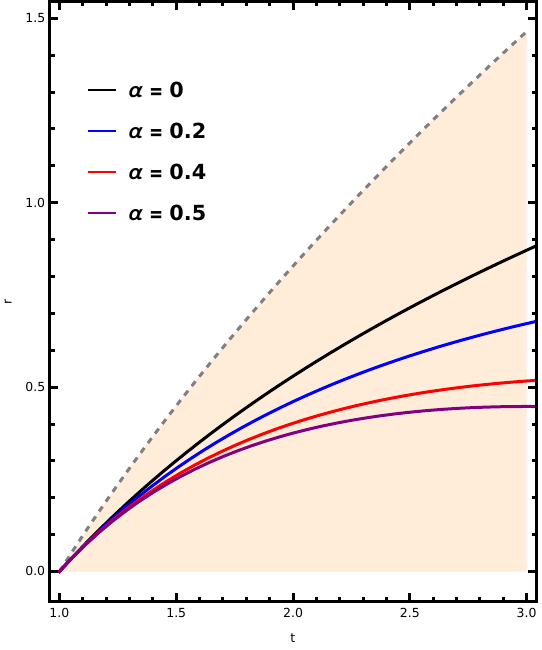} &
\includegraphics[width=0.27\textwidth]{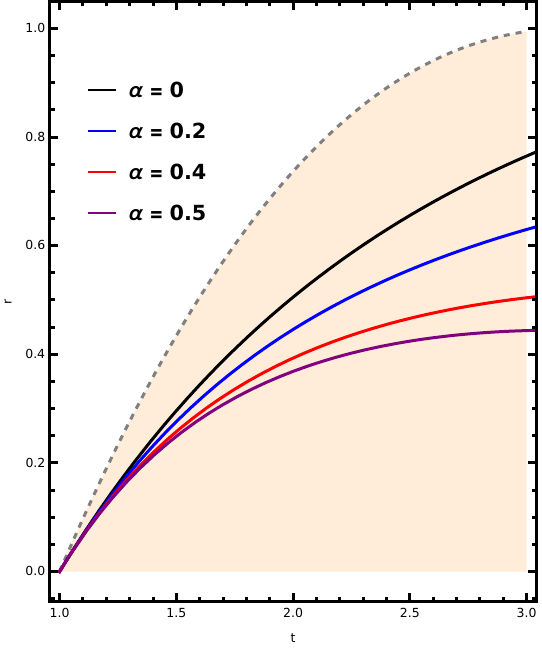} \\
\includegraphics[width=0.27\textwidth]{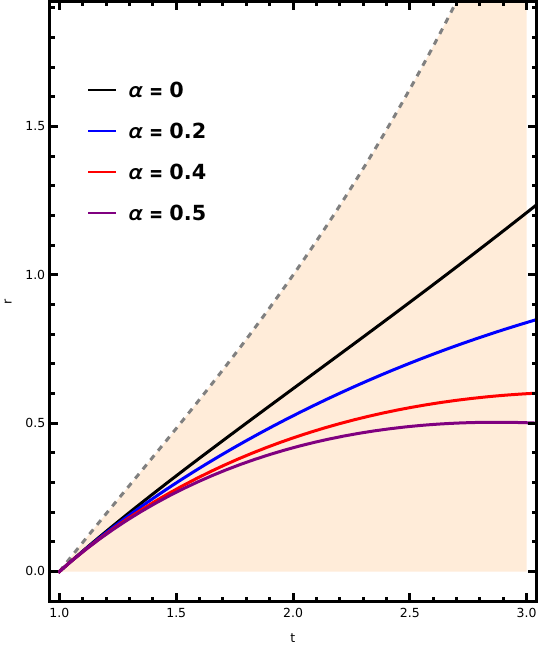} &
\includegraphics[width=0.27\textwidth]{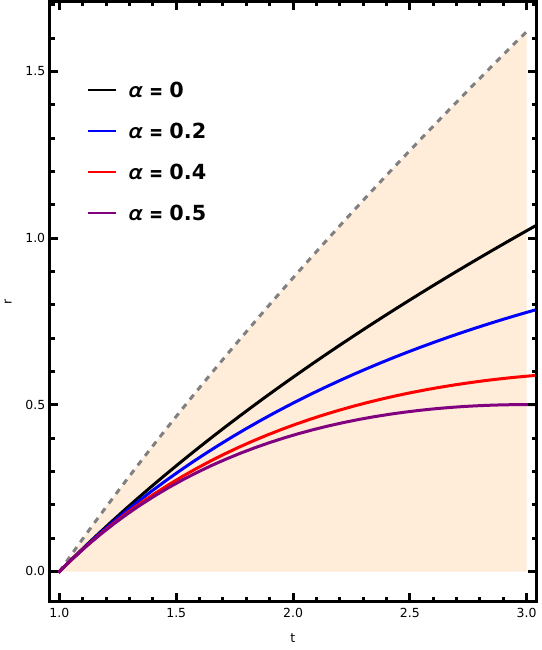} &
\includegraphics[width=0.27\textwidth]{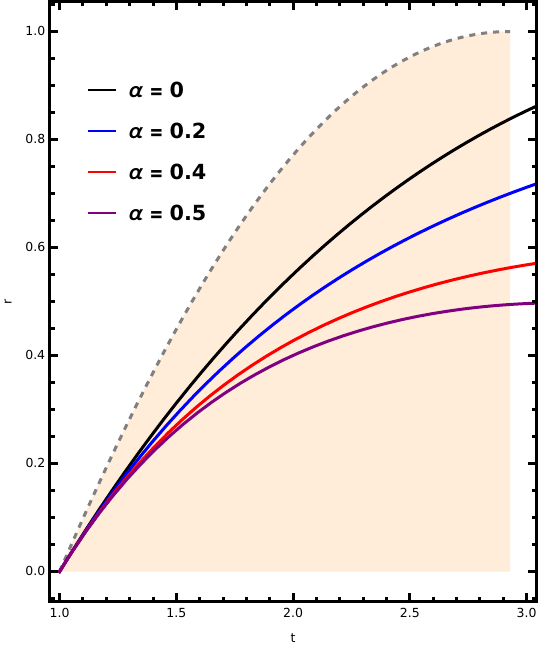} \\
\includegraphics[width=0.27\textwidth]{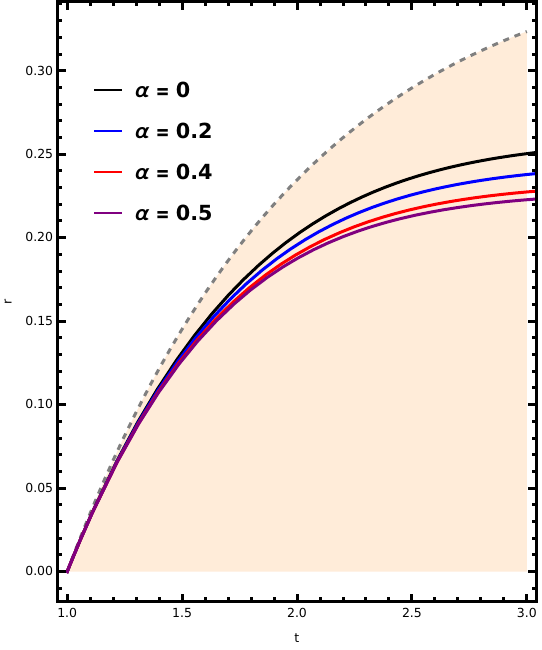} &
\includegraphics[width=0.27\textwidth]{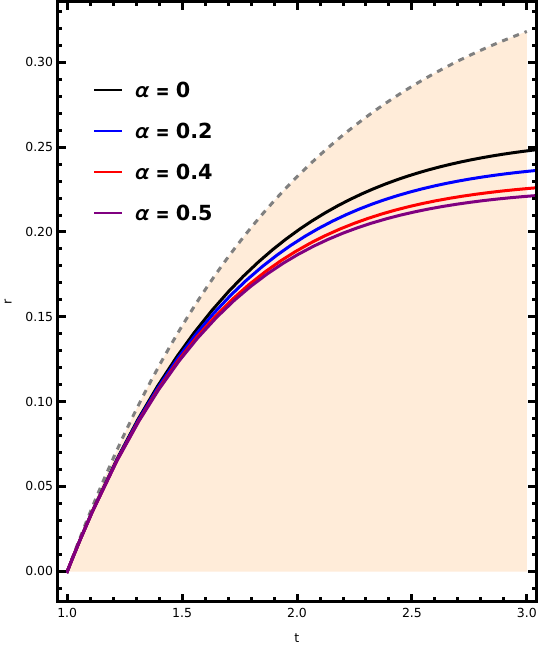} &
\includegraphics[width=0.27\textwidth]{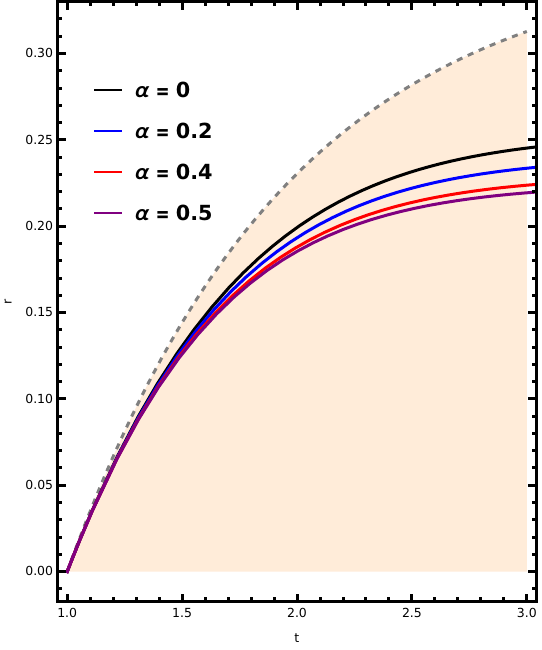} \\
\end{tabular}
\caption{Representative geodesic curves for different curvature parameter $k$ 
and different values of $\alpha$.  
Top row: Dust-dominated universe, $a(t) \propto t^{2/3}$.  
Second row: Radiation-dominated universe, $a(t) \propto t^{1/2}$.  
Third row: Stiff-fluid universe, $a(t) \propto t^{1/3}$.  
Bottom row: Dark energy-dominant universe. In each row, plots correspond to $k=-1$ (left), $k=0$ (center), and $k=1$ (right).}
\label{plot1}
\end{figure*}
The FLRW metric provides a description of a spatially homogeneous and isotropic universe in $(3+1)$ dimensions. The underlying geometry of these manifolds is characterized by a metric tensor \( g_{\mu\nu} \), defined over the spacetime indices \( \mu, \nu \in \{0,1,2,3\} \). In a coordinate system where the metric assumes a diagonal form, the infinitesimal spacetime separation between events is described by the invariant interval with signature ($-+++$) as
\begin{equation}
ds^2=-dt^2+a(t)^2\left[\frac{dr^2}{1-kr^2}+r^2(d\theta^2+\sin^2\theta d \phi^2)\right],
\end{equation}
where $a(t)$ is the scale factor encoding the dynamical evolution of spatial hypersurfaces, and $k \in \{-1, 0, +1\}$ is the curvature index corresponding to open (hyperbolic), flat (Euclidean), and closed (spherical) spatial geometries, respectively. 
The expression for the scale factor $a(t)$ in the FLRW metric is obtained through the Friedmann equation, which is the differential equation for $00$-component of the Einstein equation. 
The scale factors for
manifolds that represent spacetime with Dust, Radiation, Stiff fluid and Dark energy (DE) is given as
\begin{align}
a_{\text{Dust}}(t)  &\propto t^{\frac{2}{3}}, \label{dust} \\[1.5ex]
a_{\text{Rad}}(t)   &\propto t^{\frac{1}{2}}, \label{rad} \\[1.5ex]
a_{\text{Stiff}}(t) &\propto t^{\frac{1}{3}}, \label{stiff} \\[1.5ex]
a_{\text{DE}}(t)    &\propto e^t. \label{dark}
\end{align}
To study the motion of particles and the causal structure of the spacetime, one requires the Christoffel symbols $\Gamma^\mu_{\nu\rho}$, which encode the affine connection of the manifold and enter directly into the covariant derivative and geodesic equations. For the FLRW line element, the non-vanishing Christoffel symbols are computed as
\begin{equation}\label{chri}
\begin{aligned}
\Gamma^0_{11} &= \frac{a \dot{a}}{1 - k r^2}, &
\Gamma^0_{22} &= a \dot{a} r^2, &
\Gamma^0_{33} &= a \dot{a} r^2 \sin^2 \theta, \\
\Gamma^1_{01} &= \Gamma^1_{10} = \frac{\dot{a}}{a}, &
\Gamma^1_{11} &= \frac{k r}{1 - k r^2}, &
\Gamma^1_{22} &= -r (1 - k r^2), \\
\Gamma^1_{33} &= -r (1 - k r^2) \sin^2 \theta, &
\Gamma^2_{02} &= \Gamma^2_{20} = \frac{\dot{a}}{a}, &
\Gamma^2_{12} &= \Gamma^2_{21} = \frac{1}{r}, \\
\Gamma^2_{33} &= -\sin \theta \cos \theta, &
\Gamma^3_{03} &= \Gamma^3_{30} = \frac{\dot{a}}{a}, &
\Gamma^3_{13} &= \Gamma^3_{31} = \frac{1}{r}, \\
\Gamma^3_{23} &= \Gamma^3_{32} = \cot \theta
\end{aligned}
\end{equation}
From the Christoffel symbols calculation in Eq.(\ref{chri}), one can get the four sets of differential equations written as
\begin{align}
\frac{d^2 t}{d \lambda^2} 
&+ \Gamma^0_{\sigma\nu}\, \dot{x}^\sigma \dot{x}^\nu = - g^{00} \bigl( \partial_\sigma V_0 - \partial_0 V_\sigma \bigr) \dot{x}^\sigma, \\
\frac{d^2 r}{d \lambda^2} 
&+ \Gamma^1_{\sigma\nu}\, \dot{x}^\sigma \dot{x}^\nu = - g^{11} \bigl( \partial_\sigma V_1 - \partial_1 V_\sigma \bigr) \dot{x}^\sigma, \\
\frac{d^2 \theta}{d \lambda^2} 
&+ \Gamma^2_{\sigma\nu}\, \dot{x}^\sigma \dot{x}^\nu = - g^{22} \bigl( \partial_\sigma V_2 - \partial_2 V_\sigma \bigr) \dot{x}^\sigma, \\
\frac{d^2 \phi}{d \lambda^2} 
&+ \Gamma^3_{\sigma\nu}\, \dot{x}^\sigma \dot{x}^\nu = - g^{33} \bigl( \partial_\sigma V_3 - \partial_3 V_\sigma \bigr) \dot{x}^\sigma.
\end{align}
and this gives
\begin{align}
\frac{d^2 t}{d \lambda^2} 
&+ \frac{H a(t)^2}{1 - k r^2} \left( \frac{d r}{d \lambda} \right)^2
+ a(t)^2 H r^2 
   \Bigg[ \left( \frac{d \theta}{d \lambda} \right)^2 
   \nonumber \\
&+ \sin^2 \theta \left( \frac{d \phi}{d \lambda} \right)^2 \Bigg] = \left( \partial_\sigma V_0 - \partial_0 V_\sigma \right) 
   \frac{d x^\sigma}{d \lambda}, \\
\frac{d^2 r}{d \lambda^2} 
&+ 2H \frac{dt}{d \lambda}\frac{d r}{d \lambda}
+ \frac{kr}{1-kr^2}\left(\frac{d r}{d \lambda}\right)^2 
- r(1-kr^2) \Bigg[ \left( \frac{d\theta}{d \lambda}\right)^2
\nonumber \\
&+ \sin^2 \theta \left(\frac{d\phi}{d \lambda}\right)^2 \Bigg] = -\left(\frac{1-kr^2}{a(t)^2}\right) 
   \left( \partial_\sigma V_1 - \partial_1 V_\sigma \right)
   \frac{d x^\sigma}{d \lambda}, \\
\frac{d^2 \theta}{d \lambda^2} 
&+ 2H \frac{dt}{d \lambda}\frac{d \theta}{d \lambda}
+ \frac{2}{r}\frac{d r}{d \lambda} \frac{d \theta}{d \lambda}
- \sin \theta \cos \theta \left(\frac{d \phi}{d \lambda}\right)^2 \nonumber \\
&= -\left(\frac{1}{a(t)^2 r^2}\right)
   \left( \partial_\sigma V_2 - \partial_2 V_\sigma \right)
   \frac{d x^\sigma}{d \lambda}, \\
\frac{d^2 \phi}{d \lambda^2} 
&+ 2H \frac{dt}{d \lambda}\frac{d \phi}{d \lambda}
+ \frac{2}{r}\frac{d r}{d \lambda} \frac{d \phi}{ d \lambda}
+ 2 \cot{\theta}\frac{d \theta}{d \lambda} \frac{d \phi}{d \lambda} \nonumber \\
&= -\left(\frac{1}{a(t)^2 r^2 \sin^2{\theta}}\right) 
   \left( \partial_\sigma V_3 - \partial_3 V_\sigma \right)
   \frac{d x^\sigma}{d \lambda}.
\end{align}
In the radial-temporal plane, the system of differential equations reduces to a pair of equations by assuming $\phi=\psi=constant$,
\begin{align}
&\frac{d^2t}{d \lambda^2} + \frac{Ha(t)^2}{1-kr^2}\left(\frac{dr}{d\lambda}\right)^2 = ( \partial_\sigma V_0-\partial_0 V_\sigma)\frac{d x^\sigma}{d \lambda},  \\
&\frac{d^2r}{d \lambda^2}+2H \frac{dt}{d \lambda}\frac{d r}{d \lambda}+\frac{kr}{1-kr^2}\left(\frac{d r}{d \lambda}\right)^2   \nonumber \\
&\qquad \quad \quad \quad \quad \quad = -\left(\frac{1-kr^2}{a(t)^2}\right)( \partial_\sigma V_1-\partial_1 V_\sigma)\frac{d x^\sigma}{d \lambda}
 \end{align}

\subsubsection{Example for Conservative external force}\label{sec2b}
\begin{figure*}[t]
\centering
\begin{tabular}{ccc}
\includegraphics[width=0.27\textwidth]{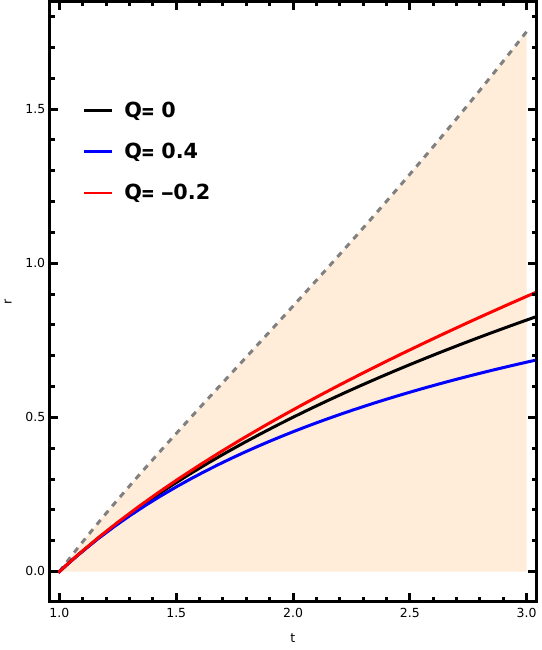} &
\includegraphics[width=0.27\textwidth]{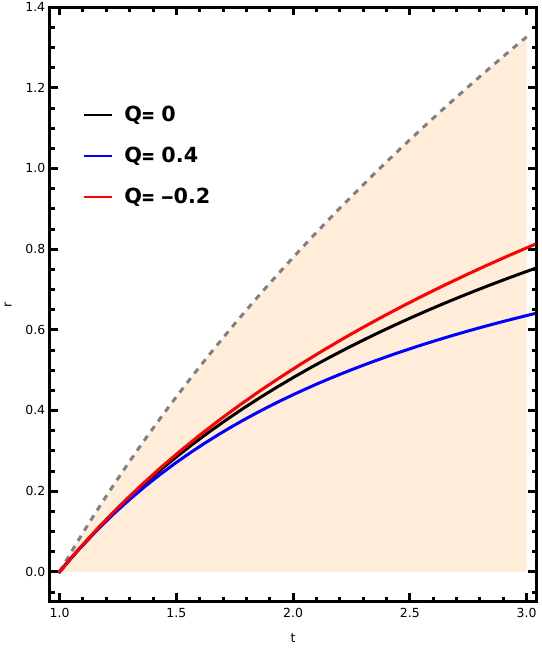} &
\includegraphics[width=0.27\textwidth]{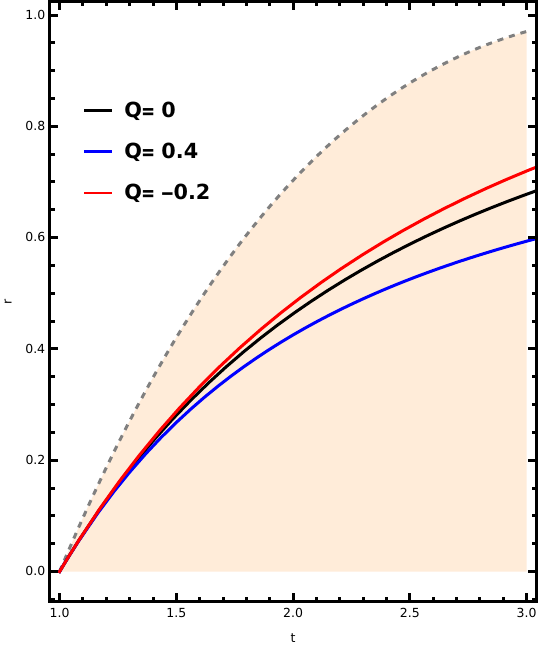} \\
\includegraphics[width=0.27\textwidth]{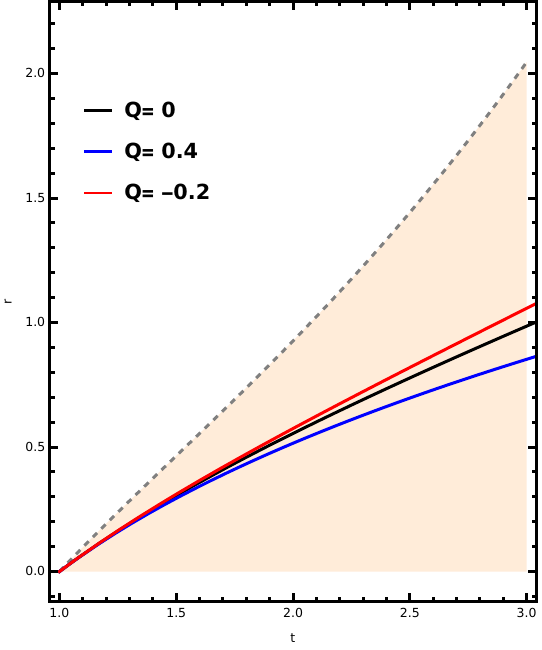} &
\includegraphics[width=0.27\textwidth]{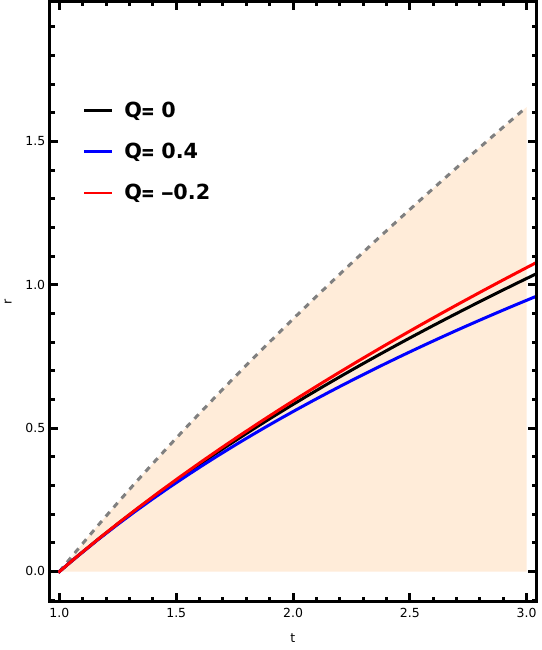} &
\includegraphics[width=0.27\textwidth]{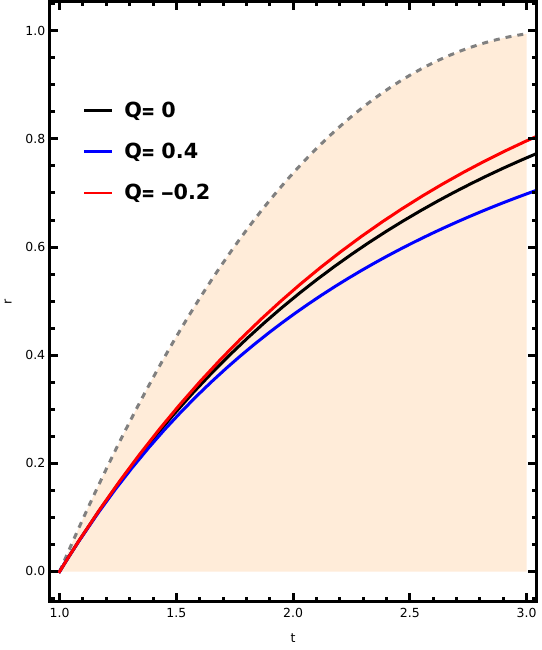} \\
\includegraphics[width=0.27\textwidth]{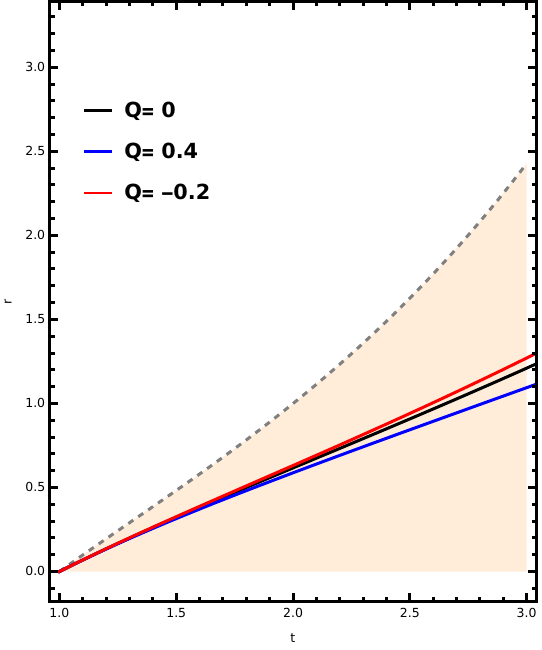} &
\includegraphics[width=0.27\textwidth]{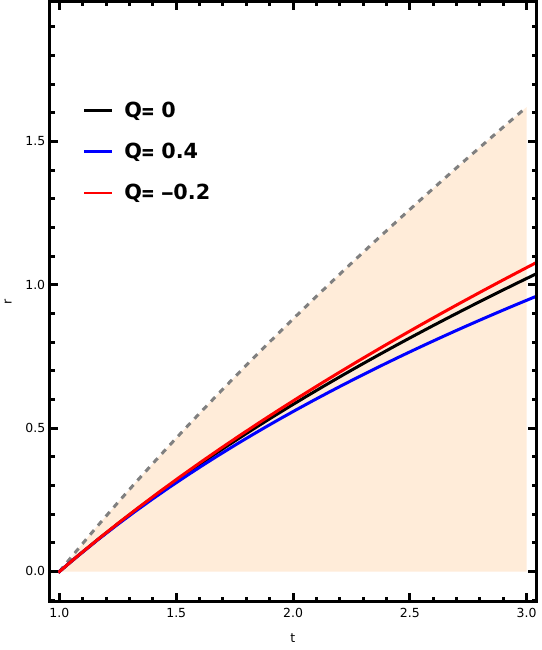} &
\includegraphics[width=0.27\textwidth]{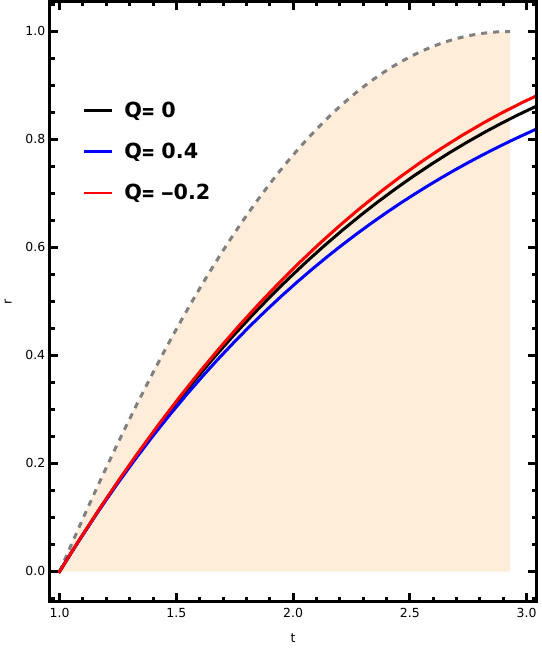} \\
\includegraphics[width=0.27\textwidth]{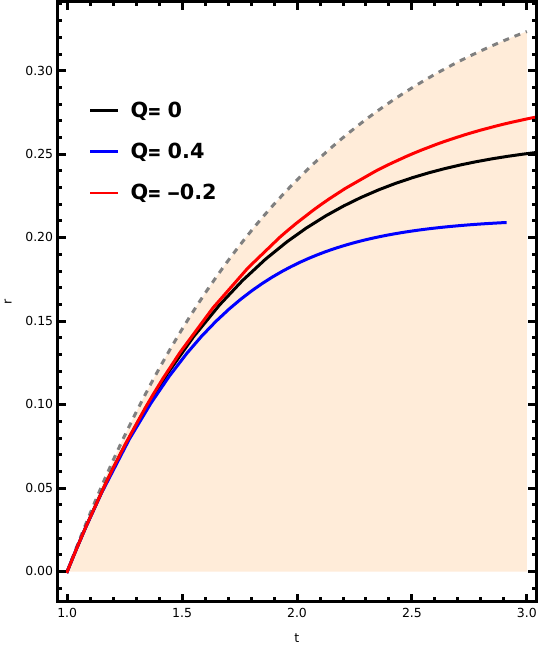} &
\includegraphics[width=0.27\textwidth]{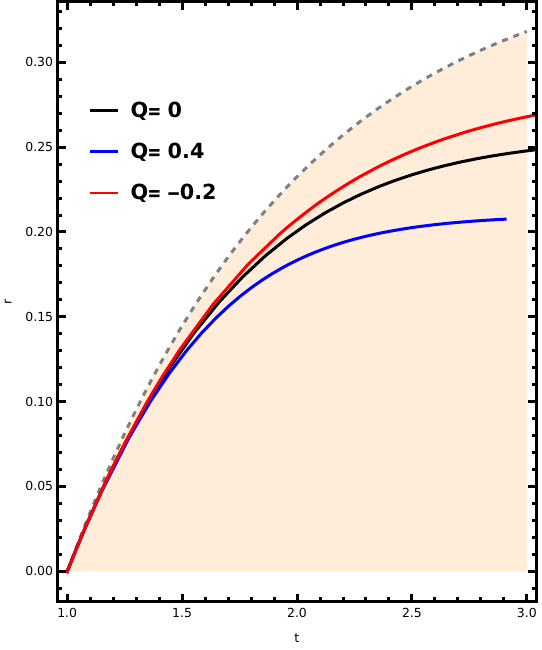} &
\includegraphics[width=0.27\textwidth]{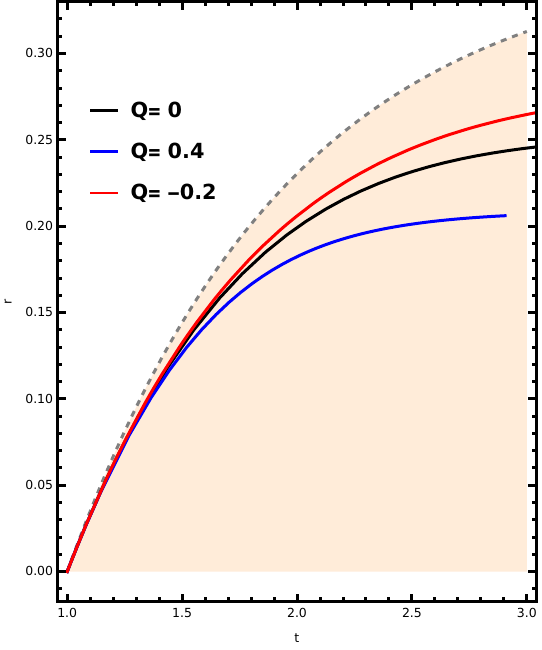} \\
\end{tabular}
\caption{The plot Represent geodesic curves for different curvature parameter $k$ 
with different dissipative term values $Q$.  
Top row: Dust-dominated universe, $a(t) \propto t^{2/3}$.  
Second row: Radiation-dominated universe, $a(t) \propto t^{1/2}$.  
Third row: Stiff-fluid universe, $a(t) \propto t^{1/3}$.  
Bottom row: Dark energy-Dominated universe. In each row, plots correspond to $k=-1$ (left), $k=0$ (center), and $k=1$ (right).}
\label{plot2}
\end{figure*}
In this small subsection, we restrict our analysis to the radial-temporal sector of the geodesic equation, fixing $\theta=\mathrm{const.}$ and $\phi=\mathrm{const.}$. We focus on the case in which the motion is influenced by a conservative external force, modeled via an electromagnetic four-potential $V_\mu$, with the associated field strength tensor 
$
F_{\mu\nu} \equiv \partial_\mu V_\nu - \partial_\nu V_\mu $.
For definiteness, we consider a purely radial electric field specified by 
$
V_0 = \alpha\, r, \quad V_1 = 0,
$
where $\alpha$ is a constant parameter that may be interpreted as the electric field strength $E$.  
Under these assumptions, the geodesic equations for the $(t,r)$ coordinates in the FRW background metric take the form
\begin{equation}
\frac{d^2t}{d \lambda^2} + \frac{Ha(t)^2}{1-kr^2}\left(\frac{dr}{d\lambda}\right)^2
= - (\partial_0 V_1- \partial_\nu V_0)\frac{d r}{d \lambda},
\label{geo_t}
\end{equation}
\begin{align}
\frac{d^2r}{d \lambda^2} + 2H \frac{dt}{d \lambda}\frac{d r}{d \lambda}
&+ \frac{kr}{1-kr^2}\left(\frac{d r}{d \lambda}\right)^2 \nonumber \\
&= \left(\frac{1-kr^2}{a(t)^2}\right)(\partial_1 V_0- \partial_0 V_1)\frac{dt}{d \lambda}.
\label{geo_r}
\end{align}
For the chosen potential $V$, the force terms on the right-hand side reduce to constants proportional to the electric field, thereby providing a direct coupling between the background expansion (through $a(t)$ and $H$) and the electromagnetic interaction. Importantly, these non-vanishing force terms possess a component parallel to the particle's 4-velocity. As a result, the qualitative shape of the geodesic trajectories remains unchanged, while only the rate of progression along the trajectory is modified as shown in Fig.(\ref{plot1}).

Figure~\ref{plot1} shows that the solutions to Eqs.~\eqref{geo_t}--\eqref{geo_r} depend sensitively on the spatial curvature parameter \(k\), the cosmological expansion era, and the external conservative force parameter \(\alpha\). Across all parameter choices, the overall trajectory shape remains qualitatively consistent. The primary effect of increasing \(\alpha\) is a systematic reduction in the particle's effective velocity along its path, manifesting as a downward deviation from the pure geodesic (\(\alpha = 0\)). Furthermore, all timelike trajectories remain confined within the causal horizon (depicted by the orange shaded region), in agreement with relativistic causality.

\subsubsection{Example for dissipative Lagrangian}\label{sec2ab}
 We turn to the question of how dissipative effects impact the geodesic motion of a test particle in an FLRW spacetime. From Eq.~(\ref{dissiex}), the dissipation exponent in the comoving frame with four-velocity $u^\kappa = (1,0,0,0)$ is given by
\begin{equation}
\Gamma = \int 3H(t)\, Q(x^\alpha)\, d\lambda,
\end{equation}
where $H(t)$ denotes the Hubble parameter.

In cosmological settings, dissipative processes can originate from a variety of microphysical and effective mechanisms, such as bulk viscosity, particle production, scalar-field interactions, and non-equilibrium thermodynamic effects. These scenarios generally lead to nontrivial functional forms of the dissipation function $Q(x^\alpha)$~\cite{A1,A2,A3,A4,A5,A7,A8,A9}. To preserve analytical tractability while isolating the qualitative impact of dissipation on geodesic evolution, we adopt the constant parametrisation, which is one of the form proposed in Ref.~\cite{R11}, $
Q(x^\alpha) = Q_0 = \text{constant}$.
This simplification maintains full compatibility with FLRW symmetry and enables a transparent analytical characterisation of the dissipative modifications to the geodesic equation.

 In this case, the geodesic equation simplifies to
\begin{equation}
\begin{split}
\frac{d^2t}{d \lambda^2} 
+ \frac{H a(t)^2}{1-kr^2}\left(\frac{dr}{d\lambda}\right)^2
= - \frac{1}{2}(\partial_0 \Gamma) \Bigg[
 \left(\frac{dt}{d\lambda}\right)^2 
\\+ \frac{a^2(t)}{1-kr^2}\left(\frac{dr}{d\lambda}\right)^2
\Bigg],
\label{Dgeo_t}
\end{split}
\end{equation}

\begin{equation}
\begin{split}
\frac{d^2r}{d \lambda^2} 
+ 2H \frac{dt}{d \lambda}\frac{dr}{d \lambda}
+ \frac{kr}{1-kr^2}\left(\frac{dr}{d \lambda}\right)^2 
= -(\partial_0\Gamma)
 \left(\frac{dt}{d \lambda}\right)\left(\frac{dr}{d \lambda}\right),
\label{Dgeo_r}
\end{split}
\end{equation}
where the term $\partial_0\Gamma = 2H(t)Q(t)$. The physical origin of dissipation in this framework is encoded in the scalar function $Q(x^\alpha)$. It acts as a phenomenological parameter that characterises the exchange of energy between the particle and an external 
environment, such as a thermal bath, background radiation, or effective degrees of freedom that have been coarse-grained out. The coupling with the spacetime 
expansion, expressed through $\nabla_\kappa u^\kappa$, ensures that the dissipative effect depends on the local geometry and flow. The geodesic path will be affected by the values of the dissipation function $Q$, which is chosen to be constant here. For positive value ($Q>0$), it means the energy is dissipating in the surrounding, while negative values of $Q$ correspond to effective energy injection as in dissipative quintessence and inflationary scenarios \cite{R11}.
The qualitative behaviour of this set of differential equations, Eq.(\ref{Dgeo_t},\ref{Dgeo_r}), through geodesic curve is represented in Fig.\ref{plot2}. We show some plot of geodesic curves for different values of $Q$.

\section{The Geodesic equations under minimal length effect (GUP)}\label{sec3}
In the absence of a complete and consistent theory of quantum gravity, a variety of distinct approaches have been developed to capture possible Planck-scale effects. Among them, the Generalized Uncertainty Principle (GUP) provides a phenomenological framework that encodes the existence of a minimal measurable length, thereby modifying the standard Heisenberg uncertainty principle. It can be expressed \cite{gup7,gup9,schimp}  as
\begin{equation}
\Delta x \,\Delta p \;\geq\; \frac{\hbar}{2} \left[ 1 + \beta \, (\Delta p)^2 \right],
\end{equation}
where $x$ and $p$ denote the position and its canonically conjugate momentum, respectively.  
The deformation parameter $\beta$ encodes the strength of quantum-gravity that becomes significant where the quantum gravitational effects are visible. The parameter $\beta$ typically arises in various approaches to quantum gravity, such as string theory, loop quantum gravity, and doubly special relativity. The dimension of $\beta$ is inverse of the square of Planck momentum as represented as $\beta = \beta_0/ (m_P c)^2$, where $m_P$ is the Planck mass and $\beta_0$ is a dimensionless parameter. The value of $\beta_0$ is model-dependent and can be fixed or constrained through phenomenological studies in both quantum-mechanical and classical regimes.

The above modified uncertainty relation is implemented at the operator level via the deformed commutator bracket 
\begin{equation}
[x_i,\,p_j] \;=\; i\hbar \,\delta_{ij} \left( 1 + \beta\, p^2 \right),
\end{equation}
where $p^2 \equiv p_k p^k$. On extending this algebra to Minkowski spacetime with Lorentzian signature $( -,+,+,+ )$ gives the relation as
\begin{equation}
[x^\mu,\,p^\nu] \;=\; i\hbar \,\eta^{\mu\nu} \left( 1 + \beta\, p^\rho p_\rho \right),
\end{equation}
where Greek indices $\mu,\nu,\rho \in \{0,1,2,3\}$ and $\eta^{\mu\nu}$ is the Minkowski metric. In the classical limit $\hbar \rightarrow 0$, the commutator is replaced by the corresponding Poisson bracket, yielding the deformed symplectic structure
\begin{equation}
\frac{1}{i\hbar} [x^\mu,\,p^\nu] \quad\longrightarrow\quad
\{x^\mu,\,p^\nu\} \;=\; \eta^{\mu\nu} \left( 1 + \beta\, p^\rho p_\rho \right).
\label{msym}
\end{equation}

It is crucial to emphasize that Eq.~\eqref{msym} does not represent an arbitrary noncanonical choice of symplectic structure.
In conventional Hamiltonian formulations, noncanonical symplectic forms may arise for example, in the quantization of dissipative systems~\cite{mont} yet such deformations are purely kinematical and can always be removed by a suitable canonical transformation, leaving the phase-space geometry globally flat.

In contrast, the deformation encoded in Eq.~\eqref{msym} originates from minimal measurable length motivated by quantum gravity phenomenology. Here, the symplectic structure depends intrinsically on the momentum coordinates, implying a nontrivial curvature of momentum space\footnote{Darboux's theorem guarantees that any symplectic manifold is locally equivalent to the canonical form; that is, locally one can always find coordinates $(q_i,p_i)$ such that $\omega = \sum_i dp_i \wedge dq_i$. This implies that classical noncanonical structures merely reflect coordinate choices rather than intrinsic curvature of phase space. The wedge product provides a coordinate-independent way to define the symplectic two-form, and for GUP-deformed systems, one must check explicitly whether $d\omega = 0$ holds \cite{kon,nair,bruno}.}. This interpretation establishes a natural link between the GUP framework and the theories of Doubly Special Relativity and relative locality, where Lorentz transformations act nonlinearly on momenta. Therefore, while the GUP algebra may resemble noncanonical formulations used in certain classical systems, their physical content is fundamentally distinct. The GUP deformation reflects an intrinsic modification of phase-space geometry, whereas classical noncanonical structures merely encode coordinate choices within a flat phase space.

It should also be noted that this procedure of taking the GUP-deformed commutator and promoting it directly to a Poisson bracket has limitations. In particular, as discussed in \cite{schimp}, when taking the formal limit 
$\hbar \rightarrow 0$, the deformation term often diverges as $(1/\hbar)$. Strictly speaking, this signals an inconsistency in deriving the classical limit directly from the quantum commutator. On the other hand, it has been argued in \cite{betasmall} that although the factor $\hbar^{-1}$ diverges in the classical limit, the GUP parameter $\beta$ encodes the strength of quantum-gravity effects, where the dimensionless parameter $\beta_0$ can be chosen to be vanishingly small. This ensures that the overall factor $\beta_0\hbar^{-1}$ remains small but finite, thereby providing a consistent classical limit.

An alternative approach is discussed in \cite{tack}, where the equivalence principle can be recovered in GUP-modified classical mechanics by considering composite bodies and postulating that the kinetic energy is additive. In this case, however, one must introduce a different deformation parameter $\beta_{0i}$ for each species $i$ of elementary particles of mass $m_i$
. Consequently, each species would be associated with a different minimal length scale $l_p=\hbar\sqrt{\beta_{0i}}$. For example, the minimal length probed by a proton would be smaller than that probed by an electron. This feature is clearly at odds with the universality of gravitation, and with the fact that the Planck length can be defined in a manner that is independent of the particle under consideration \cite{sch22}.

Nonetheless, in this work, and in line with many phenomenological studies, we adopt the modified symplectic structure \eqref{msym} as an effective description of minimal-length corrections at the classical level. 

To study the impact of GUP on particle trajectories in curved spacetime, we employ the Hamiltonian formalism with the modified symplectic structure Eq.\eqref{msym}. We analyze two cases, first, a free relativistic particle, and  
second, a particle subject to an interaction potential $V(x^\mu)$.  
In both scenarios, we derive the corresponding (possibly modified) geodesic equations and identify the terms arising purely from the minimal-length deformation.

\subsection{Free particle Geodesics}\label{sec3a}
\begin{figure*}[t]
\centering
\begin{tabular}{ccc}
\includegraphics[width=0.27\textwidth]{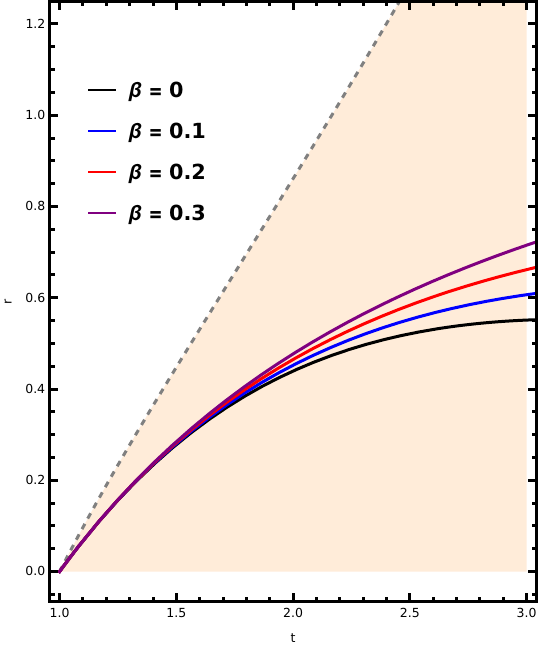} &
\includegraphics[width=0.27\textwidth]{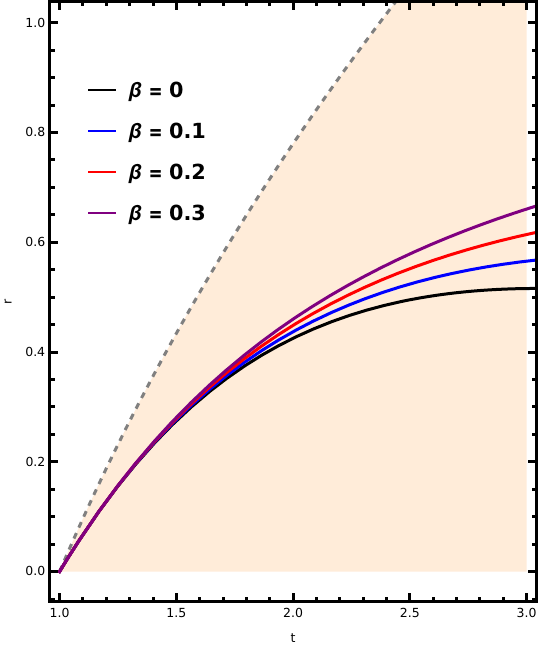} &
\includegraphics[width=0.27\textwidth]{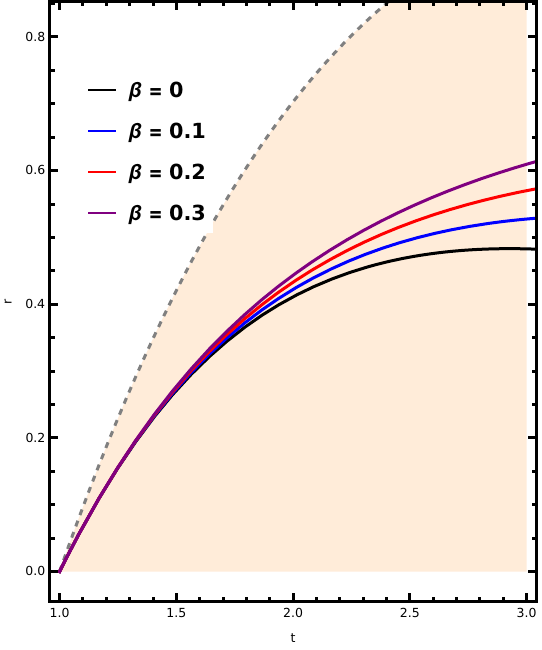} \\
\includegraphics[width=0.27\textwidth]{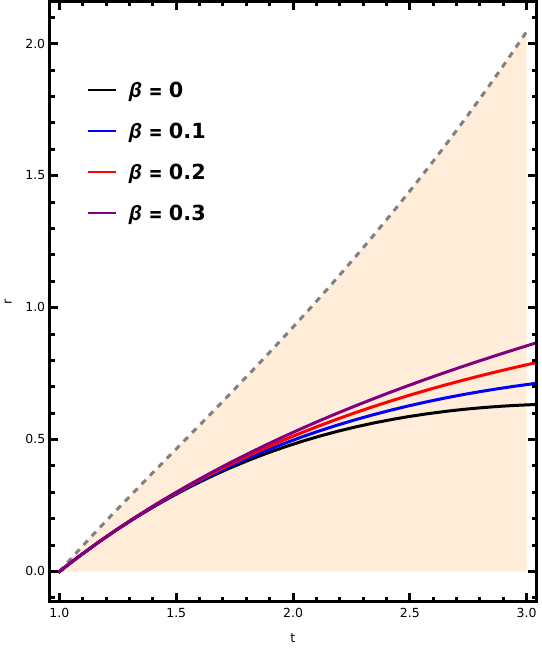} &
\includegraphics[width=0.27\textwidth]{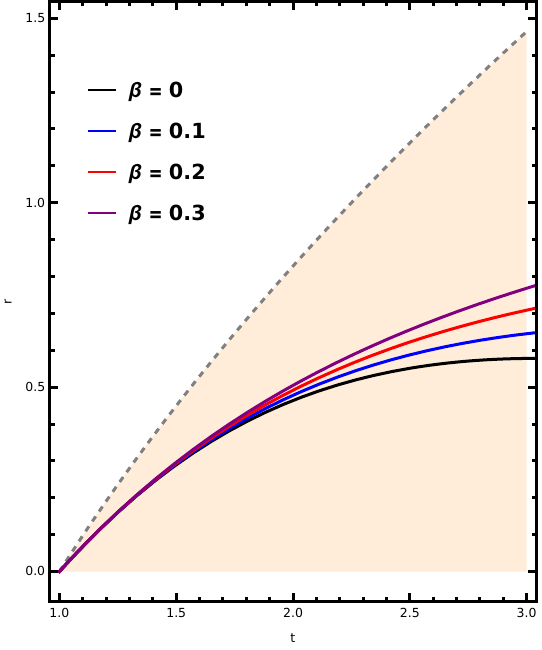} &
\includegraphics[width=0.27\textwidth]{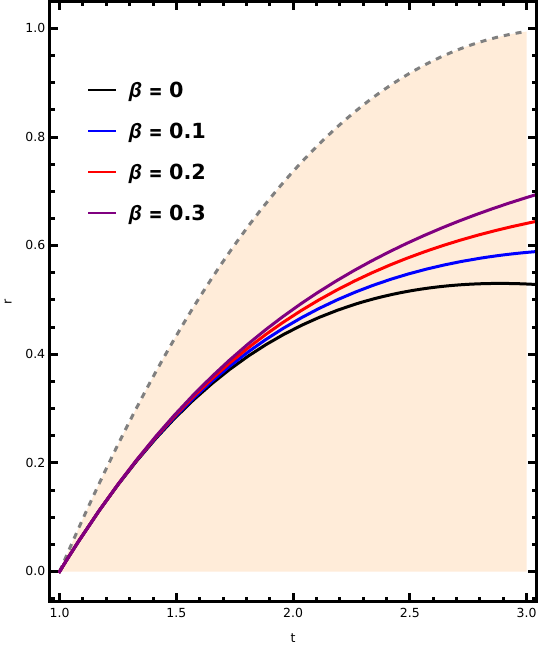} \\
\includegraphics[width=0.27\textwidth]{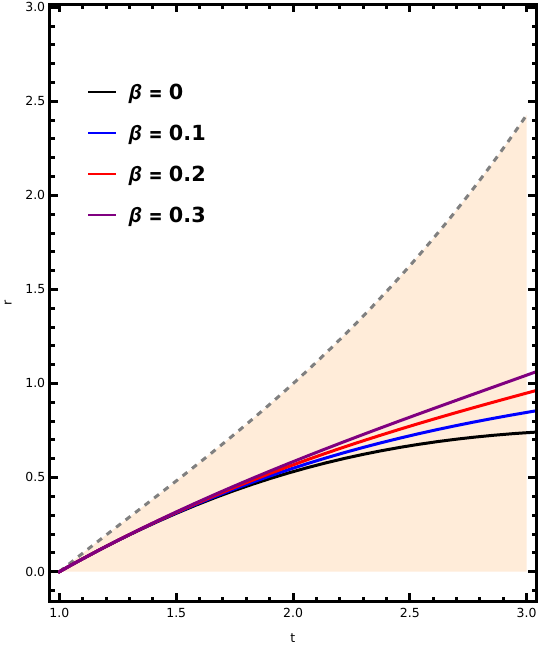} &
\includegraphics[width=0.27\textwidth]{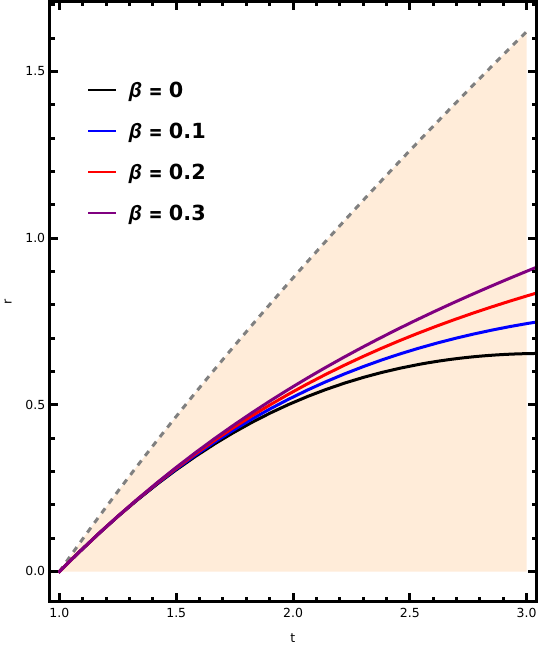} &
\includegraphics[width=0.27\textwidth]{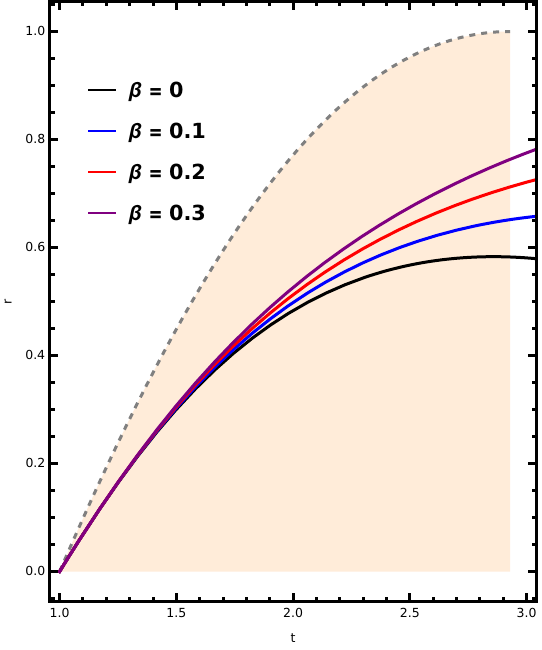} \\
\includegraphics[width=0.27\textwidth]{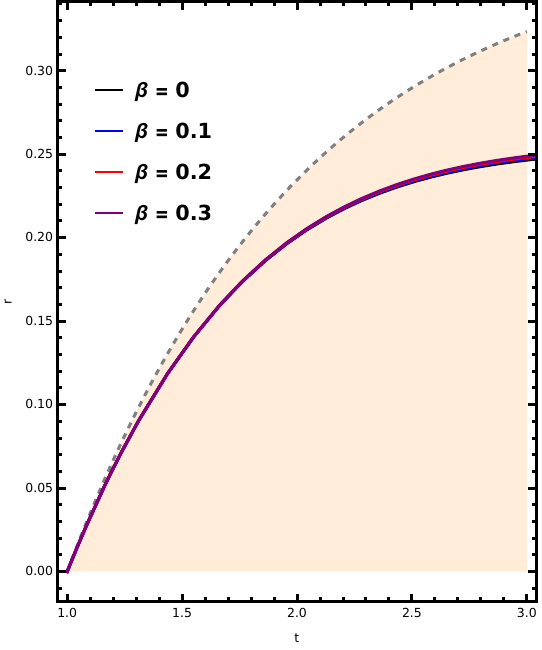} &
\includegraphics[width=0.27\textwidth]{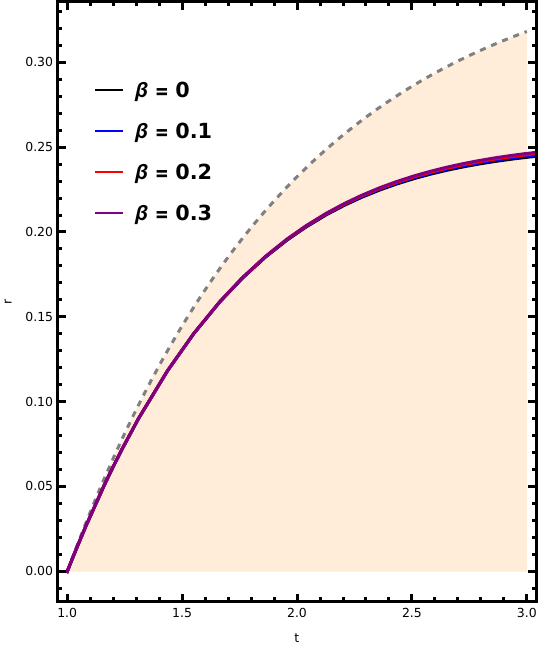} &
\includegraphics[width=0.27\textwidth]{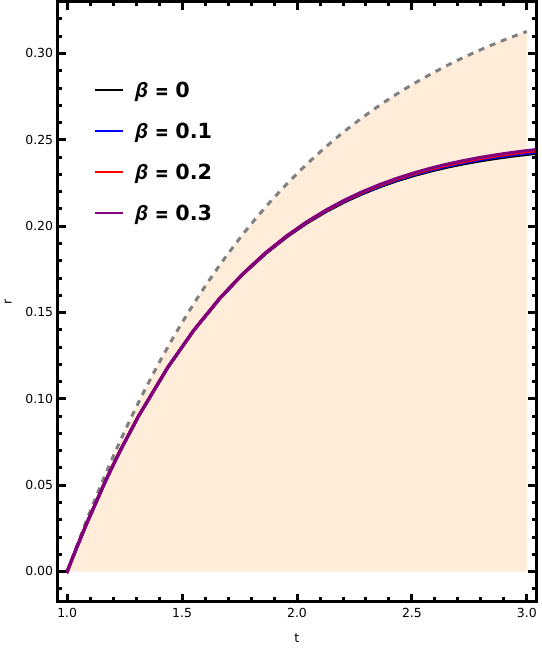} \\
\end{tabular}
\caption{Geodesic trajectories in the presence of the GUP parameter \(\beta\) for different spatial curvature values \(k = -1, 0, 1\) respectively.  
\textit{Top row:} Dust-dominated universe with scale factor \(a \propto t^{2/3}\). Increasing \(\beta\) leads to a greater deviation of the timelike geodesics from the pure geodesic, while the orange shaded region denotes the causal region.  
\textit{Second row:} Radiation-dominated universe with \(a \propto t^{1/2}\).  
\textit{Third row:} Stiff-fluid-dominated universe with \(a\propto t^{1/3}\).  
\textit{Bottom row:} Dark energy-dominated universe with \(a \propto e^{H t}\). }

\label{plot3}
\end{figure*}

We start with the free-particle Hamiltonian in curved space-time metric is represented as 
\begin{equation}
\mathcal{H}=\frac{g^{\mu\nu}}{2}p_\mu p_\nu,
\end{equation}
with the casimer operator $p^\rho p_\rho = -m^2$ for signature $-,+,+,+$. Since GUP deformed the Poisson structure, the Hamiltonian equation is written as
\begin{align}
\dot{x}^\mu &= \{x^\mu,p^\nu\}\frac{\partial \mathcal{H}}{\partial p_\nu}=g^{\mu \nu}p_\nu(1+\beta p^\rho p_\rho), \label{h1}\\
\dot{p}_\mu &= -\{x^\mu,p^\nu\}\frac{\partial \mathcal{H}}{\partial x^\mu}=-\frac{\partial_\mu g^{\alpha \beta}}{2}p_\alpha p_\beta (1+\beta p^\rho p_\rho).\label{h2}
\end{align}
Inverting the first Hamilton equation, Eq.~\eqref{h1}, to first order in $\mathcal{O}(\beta)$ and truncating higher-order terms gives the deformed momentum variable as 
\begin{equation}\label{in}
p_\nu=g_{\mu \nu} \dot{x}^\mu(1-\beta p^\rho p_\rho)+ \mathcal{O}(\beta^2) \simeq g_{\mu \nu} \dot{x}^\mu(1-\beta \dot x^\rho \dot x_\rho),
\end{equation}
now substituting $p_\nu$  from above Eq.(\ref{in}) into the second Hamilton Eq.(\ref{h2}) gives
\begin{equation}\label{h22}
\dot p_\mu = -\frac{\partial_\mu g^{\alpha \beta}}{2}(g_{\gamma \alpha} g_{\xi \beta} \dot{x}^\gamma \dot{x}^\beta)(1-\beta \dot x^\rho \dot x_\rho).
\end{equation}
Taking the derivative of Eq.(\ref{in}) w.r.t. affine parameter 
\begin{equation}\label{pdd}
    \dot{p}_\nu = (\partial_\lambda g_{\mu\nu} \dot{x}^\mu+ g_{\mu\nu}\Ddot{x}^\mu)(1-\beta \dot x^\rho \dot x_\rho)-\beta g_{\mu \nu}\dot{x}^\mu \frac{d(p^\sigma p_\sigma)}{d\lambda}+\mathcal{O}(\beta^2),
\end{equation}
Since the Casimir equation is constant along the trajectory, the second term in Eq.~\eqref{pdd} vanishes. On equating Eqs.~\eqref{pdd} and \eqref{h22} give
 \begin{align}
(\partial_\lambda g_{\mu\nu} \dot{x}^\mu+ g_{\mu\nu}\Ddot{x}^\mu)&(1-\beta \dot x^\rho \dot x_\rho)\\ \nonumber &= -\frac{\partial_\mu g^{\alpha \beta}}{2}(g_{\gamma \alpha} g_{\xi \beta} \dot{x}^\gamma \dot{x}^\beta)(1-\beta \dot x^\rho \dot x_\rho),
\end{align}
contracting with $g^{\mu\nu}$ and   the metric derivative terms combine into the Christoffel symbols gives the final unmodified geodesic equation as 
\begin{equation}
\Ddot{x}^\mu + \Gamma^\mu_{\alpha \beta} \dot x^\alpha \dot x^\beta
=0,
\end{equation}
the GUP deformation does not modify the geodesic equation for a free particle to first order in $\beta$. This result is consistent with the expectation that the quantum gravitational modification comes into the picture only through the deformed phase space structure, i.e., through modified position and momentum variables, while we have not considered minimal length effects in the geometric sector. As a result, the geodesic path of a particle is dictated solely by the unmodified background geometry.

\subsection{GUP-Corrected Geodesic Equation in the Presence of a Potential}\label{sec3b}
In this subsection, we extend the analysis to derive the geodesic equation for a particle subjected to an external potential.
We consider a point particle propagating on a curved background with metric \( g_{\mu\nu}(x) \), subject to an external potential \( V(x) \). The Hamiltonian takes the form
\begin{equation}
\mathcal{H}=\frac{1}{2}g^{\alpha\beta}p_\alpha p_\beta+V(x).   
\end{equation}  
The equations of motion follow from Hamilton's equations, modified by the GUP-deformed Poisson brackets is written as 
\begin{align}
\dot{x}^\mu &= \{x^\mu,\mathcal{H}\} 
= \bigl(1+\beta p^\rho p_\rho \bigr)\,\frac{\partial \mathcal{H}}{\partial p_\mu} 
= \bigl(1+\beta p^\rho p_\rho \bigr)\, g^{\mu\nu} p_\nu, 
\label{eq:xdot_potential}
\\[0.5ex]
\dot{p}_\mu &= \{p_\mu,\mathcal{H}\}
= -\bigl(1+\beta p^\rho p_\rho \bigr)\,\frac{\partial \mathcal{H}}{\partial x^\mu} 
\nonumber \\[0.5ex]
&= -\bigl(1+\beta p^\rho p_\rho \bigr)\!
   \left[\tfrac{1}{2}\,\partial_\mu g^{\alpha\beta}\,p_\alpha p_\beta + \partial_\mu V(x)\right].
\label{hv2}
\end{align}
From  Eq.~\eqref{eq:xdot_potential}, inverting and writing
perturbatively to first order of \(\beta\), one obtains  
\begin{equation}\label{eq:momentum_inversion}
p_\nu = g_{\mu\nu}\,\dot{x}^\mu \bigl(1-\beta p^\rho p_\rho\bigr) + \mathcal{O}(\beta^2) \simeq g_{\mu \nu} \dot{x}^\mu(1-\beta \dot x^\rho \dot x_\rho),
\end{equation}
the deformation of the phase space structure induced by the minimal length leads to a velocity-dependent correction, such that the momentum is no longer linearly related to the velocity.
Now differentiating Eq.~\eqref{eq:momentum_inversion} with respect to the affine parameter \(\lambda\) gives  
\begin{equation} \label{dotmomentum}
\dot{p}_\nu = \bigl(\partial_\lambda g_{\mu\nu}\,\dot{x}^\mu+ g_{\mu\nu}\,\ddot{x}^\mu\bigr)(1-\beta p^\rho p_\rho)
-\beta g_{\mu \nu}\dot{x}^\mu \frac{d(p^\sigma p_\sigma)}{d\lambda}+\mathcal{O}(\beta^2).
\end{equation}
The derivative of the quadratic momentum term is evaluated as  
\begin{equation}\label{affinediff}
\frac{d(p^\sigma p_\sigma)}{d\lambda} 
= \partial_\lambda g^{\rho\sigma }p_\rho p_\sigma+2g^{\rho\sigma}p_\rho \dot{p}_\sigma 
\simeq -2\,\partial_\nu V\,\dot{x}^\nu,
\end{equation}
where we used Eq.~\eqref{hv2} upto the leading order. Substituting Eq.~\eqref{affinediff}in Eq.(\ref{dotmomentum}), and eliminating $p_\nu$ through Eq.~\eqref{eq:momentum_inversion}, we obtain to order \(\mathcal{O}(\beta)\):  
\begin{equation}
\dot{p}_\nu = \Bigl(\partial_\lambda g_{\mu\nu}\,\dot{x}^\mu + g_{\mu\nu}\,\ddot{x}^\mu\Bigr)
\bigl(1 - \beta \dot{x}^\rho \dot{x}_\rho \bigr)
+ 2\beta g_{\mu\nu}\,\dot{x}^\mu\, \partial_\nu V \dot{x}^\nu 
+ \mathcal{O}(\beta^2).
\label{eq:pdot_from_x}
\end{equation}

From  Eq.\eqref{hv2} and  \eqref{eq:pdot_from_x}, and keeping terms up to \(\mathcal{O}(\beta)\), one finds
\begin{align}
g_{\mu\nu}\,\ddot{x}^\mu + \partial_\lambda g_{\mu\nu}\,\dot{x}^\mu
= &-\,\tfrac{1}{2}\,\partial_\nu g^{\alpha\beta}\, g_{\alpha\gamma} g_{\beta\delta}\,\dot{x}^\gamma \dot{x}^\delta
\\ \nonumber &- \partial_\nu V(x) \bigl(1 +2\beta\,\dot{x}^\rho \dot{x}_\rho \bigr).
\end{align}
Raising the index with \(g^{\nu\sigma}\) and using the definition of the Christoffel symbol
\[
\Gamma^\sigma_{\alpha\beta} = \tfrac{1}{2}\,g^{\sigma\nu}\,
\bigl(\partial_\alpha g_{\nu\beta} + \partial_\beta g_{\nu\alpha} - \partial_\nu g_{\alpha\beta}\bigr),
\]
the modified geodesic equation becomes
\begin{equation}
\ddot{x}^\mu + \Gamma^\mu_{\alpha\beta}\,\dot{x}^\alpha \dot{x}^\beta
= -\,g^{\mu\nu}\,\partial_\nu V(x)\,
\Bigl(1 + 2\beta\,\dot{x}^\rho \dot{x}_\rho\Bigr) + \mathcal{O}(\beta^2).
\label{eq:geodesic_GUP_potential}
\end{equation}
Equation~\eqref{eq:geodesic_GUP_potential} shows that the GUP correction appears as a velocity-dependent renormalization of the potential gradient. 
Importantly, the velocity dependence in the geodesic equation indicates a violation of the equivalence principle. In the classical regime where $\beta=0$, we recovered the standard geodesic equation and the equivalence principle.  As discussed, deriving such corrections solely through deformed Poisson brackets corresponds to considering no quantum corrections to the background metric. We emphasize that this conclusion holds within our approximation of deformed Poisson brackets on a fixed classical background, a more complete treatment should also consider possible GUP-induced corrections to the background metric. 

To illustrate the effect of the deformation on bound motion, we also examined a quadratic confining potential, $V(r)=\epsilon r^2$, for which the geodesic trajectories acquire characteristic $\beta$-dependent deviations. Representative trajectories for different values of $\beta$ are shown in Fig.~\ref{plot3}. 

\section{conclusion}\label{sec4}

In this work, we have derived and analyzed generalized geodesic equations in curved spacetime by incorporating conservative forces, dissipative effects, and quantum-gravity inspired corrections arising from the Generalized Uncertainty Principle (GUP). Starting from a variational principle, we showed that conservative forces enter the geodesic equation through antisymmetric field-strength terms, reproducing the covariant Lorentz force as a specific case. Similarly, dissipative dynamics were captured through an exponential rescaling $e^{\Gamma(g_{\alpha \beta},x^\alpha)}$ of the Lagrangian, leading to modified equations with explicit drag and gradient-like contributions. Such dissipative effects can naturally arise, for example, through couplings between scalar fields and the spacetime background. More generally, the presence of any external force can be encoded on the right-hand side of the geodesic equation as $u^\nu \nabla_\nu u^\mu = \widetilde{Q}$,
where $\widetilde{Q}$ denotes contributions from non-gravitational forces beyond the purely geometric effects of spacetime curvature.  

We further extended our analysis by incorporating minimal-length effects through a GUP-deformed symplectic structure at the classical level. Our results show that free-particle geodesics remain unaffected to leading order in the deformation parameter $\beta$, thereby preserving the equivalence principle in the absence of external interactions. However, when external potentials are present, the geodesic equation acquires velocity-dependent corrections proportional to $\beta$. These velocity-dependent modifications in the GUP-modified geodesic equation signal violations of the equivalence principle.

A noncanonical symplectic structure alone is insufficient to uniquely attribute the behavior to quantum-gravity effects. Noncanonical phase-space structures arise naturally in purely classical systems for example, in the constrained Hamiltonian dynamics \cite{mont} or in the Hamiltonian description of charged particles in electromagnetic fields \cite{Sch}, in such cases they simply reflect nontrivial coordinate choices on an otherwise flat phase space. By contrast, the GUP-deformed symplectic structure employed in this work is motivated by quantum-gravity phenomenology as it encodes a minimal measurable length and leads to a momentum-dependent deformation of the underlying phase space. Thus, while classical and GUP-induced noncanonical structures may appear algebraically similar, their physical interpretations are fundamentally distinct.

Altogether these findings demonstrate that while pure geodesic motion is robust against Planck-scale modifications, the interplay between external forces, dissipation, and GUP corrections leads to a rich phenomenology, particularly in cosmological backgrounds. Our qualitative analysis of trajectories in dust, radiation, stiff matter, and dark-energy dominated universes illustrates how such effects can alter the geodesic paths at both the classical and quantum-gravity levels, as depicted in Figs.~(\ref{plot1}, \ref{plot2}, \ref{plot3}).  

While we adopt a constant dissipation coefficient $Q_0$ to obtain a qualitative analysis, the present framework can be systematically extended to more general forms of $Q(x^\alpha)$. Adopting dissipation functions parametrized by the Hubble rate, the matter density, or the redshift would facilitate an examination of how distinct dissipative mechanisms affect geodesic trajectories and geodesic deviation in FLRW spacetimes, contributing to a broader phenomenological understanding of dissipative gravitational dynamics.

An important limitation of the present study is that GUP corrections were implemented solely through phase-space deformations, while the background metric itself was kept classical. A more complete framework should also account for GUP-induced modifications of the spacetime geometry, as suggested in studies of black-hole thermodynamics and deformed Schwarzschild metrics. Future work will therefore aim to extend this formalism to include metric deformations, composite-body effects, and observable consequences in astrophysical and cosmological settings, such as perihelion precession, gravitational lensing, and light deflection.  

In summary, our results contribute to the ongoing effort to understand how quantum-gravity motivated minimal-length effects manifest in geodesic motion, providing a bridge between classical relativity, modified dynamics, and phenomenological quantum gravity.  
\section*{Appendix A: Derivation of the Generalized Uncertainty Principle}
\label{app:GUP}

In this Appendix we derive the commonly used form of the Generalized Uncertainty Principle (GUP)
starting from the modified canonical commutator
\begin{equation}
[x,p]=i\hbar\,[\,1+\beta p^2\,],
\label{modcomm}
\end{equation}
which appears in several phenomenological models of Planck-scale physics \cite{gup7,gup9,schimp}. The Robertson (or Robertson-Schr\"odinger) uncertainty relation for two self-adjoint operators \(A\) and \(B\) states
\begin{equation}
\Delta A\,\Delta B \;\ge\; \frac{1}{2}\bigl|\langle [A,B]\rangle\bigr|,
\label{robertson}
\end{equation}
considering the operators as \(A=x\) and \(B=p\) and using Eqns.(\ref{modcomm},\ref{robertson}) gives
\begin{equation}
\Delta x\,\Delta p \;\ge\; \frac{1}{2}\bigl|\langle[x,p]\rangle\bigr|
= \frac{\hbar}{2}\bigl\langle 1+\beta p^2\bigr\rangle
= \frac{\hbar}{2}\bigl(1+\beta\langle p^2\rangle\bigr).
\label{step1}
\end{equation}

Now express \(\langle p^2\rangle\) in terms of the momentum variance and mean,
\begin{equation}
\langle p^2\rangle=(\Delta p)^2+\langle p\rangle^2 \;\ge\; (\Delta p)^2,
\end{equation}
so Eq.(\ref{step1}) implies the commonly quoted GUP bound
\begin{equation}
\;\Delta x\,\Delta p \;\ge\; \frac{\hbar}{2}\bigl(1+\beta(\Delta p)^2\bigr)\; .
\label{GUP}
\end{equation}

Equation (\ref{GUP}) makes explicit that large \(\Delta p\) increases the lower bound on \(\Delta x\),
leading to a nonzero minimum position uncertainty. To find the minimal \(\Delta x\), rewrite the Eq.(\ref{GUP}) as
\[
\Delta x \;\ge\; \frac{\hbar}{2}\Bigl(\frac{1}{\Delta p} + \beta\,\Delta p\Bigr),
\]
and minimize the right-hand side with respect to the positive variable \(\Delta p\). The derivative
vanishes at 
\[
-\frac{1}{(\Delta p)^{2}} + \beta = 0
\quad \Rightarrow \quad
\Delta p_{\rm crit} = \frac{1}{\sqrt{\beta}}.
\]
Substituting back gives the minimal uncertainty in  position as
\begin{equation}
\Delta x_{\rm min} 
= \hbar\sqrt{\beta}.
\label{xmin}
\end{equation}
Thus, the modified commutation relation (\ref{modcomm}) implies the existence of a minimal measurable length \(\Delta x_{\rm min}=\hbar\sqrt{\beta}\).

\begin{acknowledgments}
Part of this work was carried out during G.B.'s visit to IUCAA, Pune, and G.B. thanks Prof. Ranjeev Mishra for support during the stay. SDP acknowledges the indirect support of the global community of researchers, institutions, and organizations dedicated to advancing fundamental physics, which makes work of this kind possible. The work of M.K. was performed in Southern Federal University with financial support of grant of Russian Science Foundation № 25-07-IF.
\end{acknowledgments}

\end{document}